\documentstyle[floats,prd,aps,epsf]{revtex}

\draft
\begin{document}

\def\agt{\mathrel{\raise.3ex\hbox{$>$}\mkern-14mu\lower0.6ex\hbox{$\sim$}}}
\def\alt{\mathrel{\raise.3ex\hbox{$<$}\mkern-14mu\lower0.6ex\hbox{$\sim$}}}

\newcommand{\beq}{\begin{equation}}
\newcommand{\eeq}{\end{equation}}
\newcommand{\beqn}{\begin{eqnarray}}
\newcommand{\eeqn}{\end{eqnarray}}
\newcommand{\pa}{\partial}
\newcommand{\vp}{\varphi}
\newcommand{\varep}{\varepsilon}
\newcommand{\ep}{\epsilon}
\newcommand{\comp}{(M/R)_\infty} 
\def\bI{\hbox{$\,I\!\!\!$--}}

\twocolumn[\hsize\textwidth\columnwidth\hsize\csname
@twocolumnfalse\endcsname

\begin{center}
{\large\bf{Constraining nuclear 
equations of state using gravitational waves from hypermassive 
neutron stars}}
~~\\
~~\\
Masaru Shibata
~~\\
~~\\
{\em Graduate School of Arts and Sciences, 
University of Tokyo, Komaba, Meguro, Tokyo 153-8902, Japan 
}
\end{center}

\begin{abstract}
Latest general relativistic simulations for merger of binary neutron stars
with realistic equations of state (EOSs) show that a hypermassive neutron star
of an ellipsoidal figure is formed after the merger if the
total mass is smaller than a threshold value which depends on
the EOSs. The effective amplitude of
quasiperiodic gravitational waves from such hypermassive neutron stars is 
$\sim 6$--$7 \times 10^{-21}$ at a
distance of $50$ Mpc , which may be large enough for detection
by advanced laser interferometric gravitational wave detectors
although the frequency is high $\sim 3$ kHz.
We point out that the detection of such signal may lead to constraining 
the EOSs for neutron stars. 
\end{abstract}
\pacs{04.25.Dm, 04.30.-w, 04.40.Dg}
\vskip2pc]

\section{Introduction}

Binary neutron stars \cite{HT,Stairs} inspiral as a
result of the radiation reaction of gravitational waves, and
eventually merge. The most optimistic scenario based mainly on a
recent discovery of binary system PSRJ0737-3039 \cite{NEW} suggests
that such mergers may occur approximately once per year within a
distance of $\sim 50$ Mpc \cite{BNST}. Even the most conservative
scenario predicts an event rate approximately once per year within a
distance of $\sim 100$ Mpc \cite{BNST}.
This indicates that the detection rate of gravitational waves by the
advanced laser interferometric detectors such as advanced LIGO
will be $\sim 40$--600 yr$^{-1}$ \cite{BNST}, and hence, 
the merger of binary neutron stars is one of the most promising
sources for them \cite{KIP}.
Since the detection rate is likely to be very high, 
a detailed study for binary neutron stars and for 
the property of neutron stars will be possible by the data analysis 
of gravitational waves. 

Gravitational waves in the coalescing binaries will be primarily detected for
the inspiraling phase in which two stars adiabatically approach 
due to gravitational wave emission. 
In this phase, the emission time scale of gravitational waves is
much longer than the orbital period, and hence, 
the frequency and amplitude of gravitational waves increase 
in the time scale much longer than the orbital period, resulting in 
the emission of the so-called chirp signal.
The theoretical templates for the chirp signal have been computed in 
the post Newtonian theory with a very high accuracy \cite{blanchet}. 
They will be used for the matched filtering technique in the
data analysis for the detection and possibly 
for determining the mass and the spin of binary components \cite{CF}.
In the following argument, we assume that mass of two neutron stars
will be determined from the chirp signal since we consider the
events of the distance smaller than $\sim 100$ Mpc and, hence,
the signal to noise ratio is likely to be high enough for the
determination. 

When the orbital separation decreases to $\sim 3R$ where $R$
denotes the radius of neutron stars $\sim 10$--15 km, 
the merger will set in \cite{USE,T}. 
At the onset of the merger, the nature 
of the gravitational waveforms changes from the chirp-type signal to the 
burst-type one. The frequency of gravitational waves at the 
transition is likely to be $f_{\rm tran} \sim [GM/(3R)^3]^{1/2}/\pi
\approx 0.9 \pm 0.3$ kHz for $M = 2.8 M_{\odot}$. Here, the 
precise value of $f_{\rm tran}$ 
depends sensitively on the radius of neutron stars. 
Thus, if it is determined by the detection of gravitational waves, 
the equations of state (EOSs) for neutron stars may be constrained \cite{C}.
This stimulates detailed numerical simulations at the transition phase
in post Newtonian gravity (e.g., \cite{Centrella,FR}).

As mentioned above, the gravitational waveforms in
the merger phase is likely to depend sensitively on the intrinsic
property of neutron stars such as the mass, the radius, and the EOSs. 
In this letter, we focus on gravitational waves from
hypermassive neutron stars (HMNSs) which will be formed after
the merger of relatively small total mass, 
and propose a method for constraining the EOSs 
using such gravitational waves. To explain the method 
in the following, we assume that the 
mass ratio is close to unity since observed 
binary neutron stars for which the mass is determined accurately 
have such mass ratios \cite{Stairs} (cf. Table I). 

\begin{table}[b]
\begin{center}
{\footnotesize 
\begin{tabular}{ccc} \hline
PSR & $M$ & Mass ratio \\ \hline \hline
B1913+16  & 2.828 & 0.963\\ \hline
B1534+12  & 2.678 & 0.991\\ \hline
B2127+11C & 2.71  & 0.99 \\ \hline
J0737-3039 & 2.59 & 0.94 \\ \hline
\end{tabular}
}
\caption{The total mass and the mass ratio 
of observed binary neutron stars for which each mass is determined
accurately. The data are quoted from [2]. 
}
\end{center}
\end{table}

After the merger sets in, the massive merged object collapses to
a black hole or settles down to a HMNS 
depending mainly on the total mass of binaries \cite{STU,STU2}.
Here, we note that the 
``black hole formation'' is referred to as the case in which
a black hole is formed promptly after the onset of the merger.
The HMNS is defined as 
a differentially rotating neutron star for which the total
baryon rest-mass is larger than the 
maximum allowed value of rigidly rotating neutron stars for a
given EOS \cite{BSS}.

To theoretically clarify the outcome, 
fully general relativistic simulation is the unique approach. 
Over the last few years, numerical methods for solving coupled 
equations of the Einstein and hydrodynamic equations have been 
developed for this issue. Now such simulations are feasible with an
accuracy high enough for yielding scientific results (e.g., \cite{STU}) 
that can be used for comparison with observational data. 

An important finding in the latest general relativistic simulations
with realistic EOSs \cite{STU2} is that the threshold
mass for the prompt black hole formation, $M_{\rm thr}$, depends
sensitively on the EOSs.  In \cite{STU2}, the SLy and
FPS EOSs \cite{PR,DH} are used with a correction
for the thermal pressure that plays a role in the merger in which
shocks are generated (see \cite{STU2} for detail). 
In these EOSs, the values of
$M_{\rm thr}$ are $\sim 2.7M_{\odot}$ and $\sim 2.5M_{\odot}$
for SLy and FPS EOS, respectively.
Here, we should note that the maximum mass for spherical
neutron stars in the SLy and FPS EOSs are $M_{\rm
max:sph} \approx 2.04 M_{\odot}$ and $1.80M_{\odot}$, respectively.
Thus, $M_{\rm thr}$ is much larger than $M_{\rm max:sph}$. This is due
to the fact that the merged object has a large angular momentum which
results from the orbital angular momentum, and hence, the large
centrifugal force becomes available for sustaining the large
self-gravity to yield a HMNS \cite{BSS}.

Also, interesting is that the theoretical value of $M_{\rm thr}$
for the realistic EOSs is close to the mass
of observed binary systems in nature \cite{Stairs} (cf. Table I).
This suggests that the value of $M_{\rm thr}$ may be
determined from certain observational results, and can be used for 
constraining the EOSs. Thus, in this letter, we propose
a method for constraining $M_{\rm thr}$ using the signal of gravitational 
waves emitted from HMNSs formed after the merger
of binary neutron stars.

\begin{figure}[t]
\vspace{-4mm}
\begin{center}
\epsfxsize=3in
\leavevmode
\epsffile{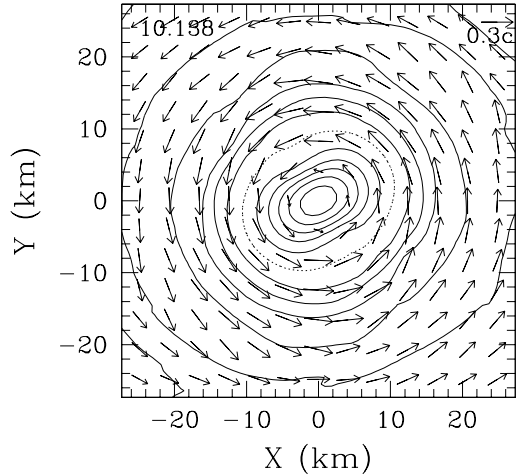}
\vspace{-6mm}
\caption{\small
The density contour curves for $\rho$ in the equatorial plane
at $t=10.138$ ms. (The initial condition is set at $t=0$ and
the merger sets in at $t \sim 2$ ms.) 
The solid contour curves are drawn for
$\rho=2\times 10^{14} \times i ~{\rm g/cm^3}~(i=2 \sim 10)$ and for 
$2\times 10^{14} \times 10^{-0.5 i}~{\rm g/cm^3}~(i=1 \sim 7)$. 
The dotted curves denote $2 \times 10^{14}~{\rm g/cm^3}$. 
Vectors indicate the local velocity field $(v^x,v^y)$, and the scale 
is shown in the upper right-hand corner.
\label{FIG1}}
\end{center}
\end{figure}

\section{Method}

For binaries of mass larger than $M_{\rm thr}$,
most of the fluid elements with 
more than 99\% of the total mass collapse to a black hole directly
in the merger \cite{STU2}.
In such case, gravitational waves associated with the quasinormal mode 
ringing of the formed black hole will be emitted. The latest general
relativistic simulations \cite{STU,STU2} have shown that the 
nondimensional spin parameter $a \equiv cJ/GM^2$ for the formed black hole
is $\sim 0.7$--0.8. Here $M$, $J$, $G$, and $c$ denote the 
mass and angular momentum of the black hole, the gravitational constant, and
the speed of light, respectively. This suggests that the
frequency of gravitational waves will be very high as
6.5--7$(2.8M_{\odot}/M)$ kHz \cite{leaver}. This value is far out of the
best sensitive frequency range of the laser interferometric
gravitational wave detectors \cite{KIP}. 
This implies that the Fourier spectrum of 
gravitational waves will not have any peak for the frequency between
1 and $\sim 6$ kHz for the case of prompt black hole formation. 

A HMNS is formed after the merger temporarily for $M < M_{\rm thr}$. 
General relativistic simulations with the SLy and FPS EOSs
have clarified \cite{STU2} that the HMNS 
has a highly ellipsoidal shape (see Fig. 1). Such ellipsoidal shape
results from the fact that the HMNS is rapidly
rotating and the adiabatic index of the EOSs is very large
($\agt 2.5$) \cite{PR,DH}. 

\begin{figure}[t]
\vspace{-4mm}
\begin{center}
\epsfxsize=3.in
\leavevmode
\epsffile{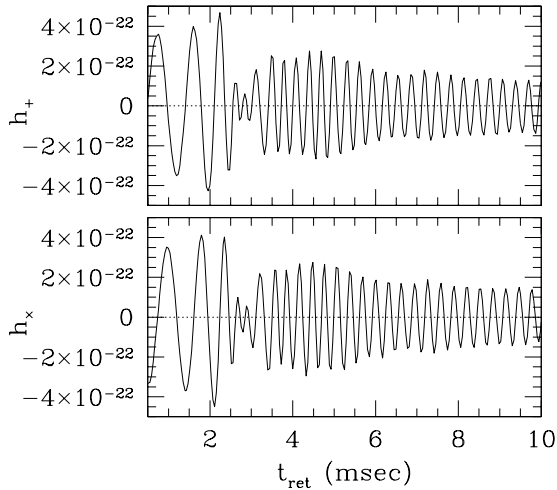}
\end{center}
\vspace{-6mm}
\caption{$h_+$ and $h_{\times}$ modes of gravitational waves
observed along the rotational axis of the binary neutron stars
at a hypothetical distance of 50 Mpc. 
Mass of two neutron stars is identical and $1.3M_{\odot}$. 
The SLy EOS is adopted. 
\label{FIG2} }
\end{figure}

\begin{figure}[thb]
\vspace{-4mm}
\begin{center}
\epsfxsize=3.in
\leavevmode
\epsffile{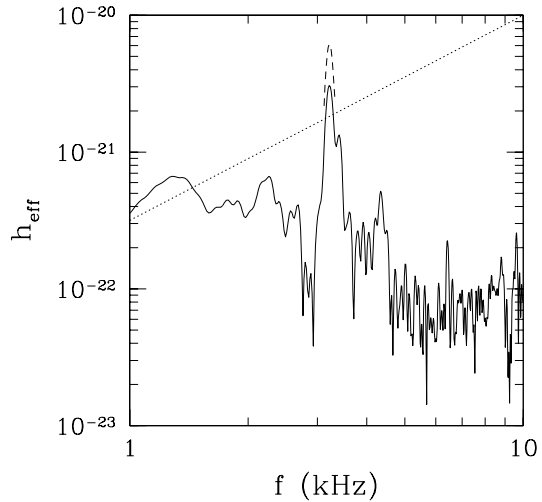}
\end{center}
\vspace{-6mm}
\caption{The (nondimensional)
effective amplitude $h_{\rm eff}(f)$ of gravitational waves
observed along the rotational axis of the binary neutron stars 
at a distance of 50 Mpc. The solid curve is the spectrum computed from
gravitational waveforms shown in Fig. 2. The dashed curve denotes 
an expected peak amplitude computed for a long term integration. 
The dotted line denotes the planned noise level of the advanced LIGO. 
\label{FIG3}}
\end{figure}

Due to the high ellipticity, the HMNS becomes a
strong emitter of quasiperiodic gravitational waves. In Fig. 2, we display 
the typical gravitational waveforms during the merger for 
the SLy EOS. In this example, two stars of the binary are 
identical and the mass of each is $1.3M_{\odot}$. The simulation was
performed with an initial condition of a quasiequilibrium circular orbit 
for which the orbital frequency is slightly smaller than that for 
the innermost stable circular orbit \cite{T} and the orbital period
is $\sim 2$ ms. In the early stage of the simulation, the binary is in a
quasistable circular orbit. Reflecting this fact, the chirp signal 
is seen for $t_{\rm ret} \alt 2$ ms. The merger begins after 
one orbit: The signal for $t_{\rm ret} \agt 2$ ms denotes 
gravitational waves emitted from the ellipsoidal HMNS 
\cite{STU2}. Figure 2 shows that waves are quasiperiodic with the 
characteristic frequency $\sim 3$ kHz and amplitude
$\sim 1.5 \times 10^{-22}$ which do not vary in a short time scale. 
This reflects the fact that the ellipsoidal figure is preserved for
a long time scale $\gg 10$ ms. 

Since gravitational waves are emitted in a quasiperiodic manner,
the signal of approximately identical frequency can be accumulated 
by a large factor. Thus, the effective amplitude $h_{\rm eff}(f)$
can be much larger than $10^{-22}$ at
a peak frequency. Here, the effective amplitude is defined by 
\beqn
h_{\rm eff}(f) \equiv
{4 \over \pi r} \sqrt{{dE\over df}}. 
\eeqn
where 
$dE/df$ denotes the energy power spectrum computed from the Fourier spectrum
of gravitational waves and is a function of $f$ (e.g., \cite{Centrella,STU2}).
$r$ denotes the distance to the source. 
In Fig. 3, we display $h_{\rm eff}$ as a function of $f$ for a 
hypothetical distance $r=50$ Mpc. 
In plotting Fig. 3, the Fourier transformation is carried out for 
$0 \alt t_{\rm ret} \alt 10$ ms. 
The dotted line in Fig. 3 is the planned (nondimensional) noise level
($h_{\rm rms}$ in the notation of \cite{KIP}) 
due to the shot noise of the laser for the advanced 
LIGO, $h_{\rm rms} \approx 10^{-21.5}(f/1~{\rm kHz})^{3/2}$ for $f \agt 1$ kHz
\cite{KIP}. Figure 3 demonstrates that quasiperiodic gravitational
waves yield a sharp peak at $f \approx 3.2$ kHz. 
Although the frequency is high and far out of the best sensitive region
for the laser interferometric detector,
the effective amplitude is larger than the noise level. 

The angular momentum of the HMNS 
is dissipated due to gravitational radiation. 
Since the self-gravity is 
sustained by a large centrifugal force, it will eventually collapse 
to a black hole after a sufficient fraction of the angular momentum
is dissipated. However, the emission time scale is not $\sim 10$ ms
but much longer, and hence,
the ellipsoidal figure is preserved to emit quasiperiodic
gravitational waves for a longer time. This indicates that 
the effective amplitude of the quasiperiodic 
waves is in reality much larger than that shown in Fig. 3. 
In \cite{STU2}, we estimated the emission time scale 
from the dissipation time scale of the angular momentum. We found that
the time scale is $\sim 30$--50 ms for the model shown in Fig. 1--3. 
This indicates that the number of the cycle for the quasiperiodic waves
is $\sim 3$--5 times more, and hence, the effective amplitude will be
by a factor of $\sim 2$ larger. In Fig. 3, we plot the plausible
effective amplitude by the dashed curve, which is twice as large as
that plotted by the solid curve. In this case, the
signal-to noise ratio (S/N) is $\sim 3$ at an event of $r=50$ Mpc
within which one event per year is expected to happen \cite{BNST}. 
We note that the simulations were also performed for binaries of mass 
$1.2$--$1.2M_{\odot}$ and $1.25$--$1.35M_{\odot}$ with the SLy
EOS, for which HMNSs are formed.
It is found that the effective amplitudes are similar to 
those presented here \cite{STU2}, and the peak frequency is in
the range between 3 and 3.5 kHz. 

The value of S/N $\sim 3$ may not be large enough for the detection
{\em in the absence of a priori information}. 
However, in the merger of binary neutron stars, the chirp signal
of gravitational waves will be detected in the inspiral phase.
Therefore, the quasiperiodic signal should be searched for
in the condition that the merger indeed happened.
Furthermore, the time of the arrival of the quasiperiodic signal is 
determined within a small uncertainty $\sim 1$ ms. These information will 
significantly improve the signal searching in the data analysis \cite{TA}. 
Thus, in the following, we assume that it will be possible to
determine if the quasiperiodic gravitational waves are present or absent. 

The absence of the signal implies that a black hole is promptly formed
after the merger, while its presence does that a HMNS 
is formed. Here, we assume that the mass of two neutron
stars in binaries will be determined from the chirp signal in the
inspiral phase for which the signal to noise ratio is
likely to be very high. Then, the presence and absence of
quasiperiodic gravitational waves can be used for determining the
threshold mass $M_{\rm thr}$ for the prompt formation of
black holes. Since the value of S/N is not likely to be very large,
the signal can be hidden in a random noise, and hence in reality, 
it may be difficult to conclude that the signal is absent. However, 
at least, its presence provides the lower
limit of $M_{\rm thr}$, and it will give a very important
information for constraining the EOSs for neutron stars.
For example, if quasiperiodic gravitational waves are detected from
a HMNS of mass $M = 2.6M_{\odot}$, the FPS EOS as well as
the EOSs with similar stiffness should be rejected
since $M_{\rm thr} \sim 2.5M_{\odot}$ for it \cite{STU2}. 
As this example shows, the advantage of this method is that 
only one detection will significantly constrain the EOSs. 
The situation is in contrast to the method in which
the value of $f_{\rm tran}$ is used to determine the radius of
neutron stars. In this method, the detailed relation between 
$f_{\rm tran}$ and $M$ is necessary for determining the
EOSs, and hence, many observational data sets are required. 

One of the remarkable findings in the latest general relativistic
simulations \cite{STU2} with realistic EOSs \cite{PR,DH} 
is that the values of $M_{\rm thr}$ 
($M_{\rm thr}\sim 2.7M_{\odot}$ and 
$\sim 2.5M_{\odot}$ for the SLy and FPS EOSs) 
are very close to the total mass of the binary neutron stars observed so far 
(cf. Table I) \cite{Stairs}. 
Namely, gravitational waves from 
the merger of two neutron stars of the total mass
$M=2.5$--$2.8M_{\odot}$ are likely to be observed frequently, and 
can be used for determining the threshold mass for the direct black
hole formation. Therefore, 
we conclude that if the sensitivity of the detectors
is improved to a planned level and if the event rate of the merger
in nature agrees with a theoretical value \cite{BNST}, 
the threshold mass will be determined 
for constraining the EOS for neutron stars. 
%
We emphasize that it is important to search for such signal 
whenever the chirp signal of gravitational waves from 
inspiraling binary neutron stars is detected and
the masses of two neutron stars are determined. 

\acknowledgments

The author thanks H. Tagoshi for explaining data analysis
techniques. 
Numerical computations were performed on the FACOM VPP5000 machines 
at the data processing center of National Astronomical Observatory
of Japan. This work was in part supported by Monbukagakusho 
Grant (Nos. 15037204, 15740142, and 16029202).

\end{document}